\documentclass[sigconf, review=false]{acmart}

\usepackage{enumitem}
\usepackage{colortbl}
\usepackage{hhline}
\usepackage{cancel}

\usepackage{makecell}
\usepackage{multirow}
\usepackage{hhline}
\usepackage{rotating}

\AtBeginDocument{%
  \providecommand\BibTeX{{%
    \normalfont B\kern-0.5em{\scshape i\kern-0.25em b}\kern-0.8em\TeX}}}

\copyrightyear{2022}
\acmYear{2022}
\setcopyright{acmcopyright}
\acmConference[ICSE-SEET '22]{44nd International Conference on Software Engineering: Software Engineering Education and Training}{May 21--29, 2022}{Pittsburgh, PA, USA}
\acmBooktitle{44nd International Conference on Software Engineering: Software Engineering Education and Training (ICSE-SEET '22), May 21--29, 2022, Pittsburgh, PA, USA}
\acmPrice{15.00}
\acmDOI{10.1145/3510456.3514152}
\acmISBN{978-1-4503-9225-9/22/05}

\begin{document}

\title{DevOps Education:\\An Interview Study of Challenges and Recommendations}

\settopmatter{authorsperrow=1} 
\newcommand{\tsc}[1]{\textsuperscript{#1}}
\author{ Marcelo Fernandes\tsc{1,2}, Samuel Ferino\tsc{1}, Anny Fernandes\tsc{1}\\
Uir{\'a} Kulesza\tsc{1}, Eduardo Aranha\tsc{1},
Christoph Treude\tsc{3} }

\affiliation{
  \institution{\vspace*{0.15cm}}
  \institution{\tsc{1}Federal University of Rio Grande do Norte (UFRN), Brazil,} \tsc{2}Federal Institute of Rio Grande do Norte (IFRN), Brazil,
  \institution{\tsc{3}University of Melbourne, Australia}
  \institution{\vspace*{0.1cm}}
  \country{}
}
\email{marcelo.fernandes@ifrn.edu.br, {samuellucas97,  anny.klarice.fernandes.127}@ufrn.edu.br}
\email{{uira, eduardoaranha}@dimap.ufrn.br, christoph.treude@unimelb.edu.au}

\renewcommand{\shortauthors}{Fernandes, et al.}

\begin{abstract}
Over the last years, the software industry has adopted several DevOps technologies related to practices such as continuous integration and continuous delivery. The high demand for DevOps practitioners requires non-trivial adjustments in traditional software engineering courses and educational methodologies. This work presents an interview study with 14 DevOps educators from different universities and countries, aiming to identify the main challenges and recommendations for DevOps teaching. Our study identified 83 challenges, 185 recommendations, and several association links and conflicts between them. Our findings can help educators plan, execute and evaluate DevOps courses. They also highlight several opportunities for researchers to propose new methods and tools for teaching DevOps.
\end{abstract}



\maketitle

\section{Introduction}

DevOps is a multidisciplinary attempt to automate the continuous delivery of new software versions while working collaboratively and guaranteeing their correctness and reliability~\cite{leite:2019}. Continuous delivery can effectively improve product and service delivery capabilities and help companies improve efficiency, thereby attracting widespread attention~\cite{yang:2020}. Recent studies show that DevOps practices are aligned with business success~\cite{chatley:2020, humble:2018}. 

As the software industry has embraced DevOps, knowledge about these practices is essential for software developers. As a consequence, the necessity for teaching DevOps has increased~\cite{perez:2021}.  However, DevOps concepts, practices, techniques and tools are not usually part of the curriculum in most educational institutions~\cite{pang:2020}. One reason for this is that teaching DevOps is challenging~\cite{fernandes:2020}. Instead of a specific technology, it involves a set of technical and non-technical concepts~\cite{alves:2021}.

Recent work~\cite{alves:2021, benni:2018, christensen:2016} proposes DevOps courses that can be used as a reference for educators who are new to this area. Recommendations and lessons learned from experienced professors can help novice professors in choosing DevOps concepts to consider and educational strategies (e.g., flipped classroom) to adopt. At the same time, challenges such as ``\textit{DevOps may be hard to understand by students who have not worked in industry or on projects of meaningful size and complexity}"~\cite{perez:2021, jones:2018} can inspire new research directions.

Fernandes et al.~\cite{fernandes:2020} characterized the area of DevOps Education by bringing together challenges and recommendations through a systematic literature review. However, their work only identified challenges and recommendations mentioned in published papers, thus missing out on challenges and recommendations that have not yet attracted the attention of academia. Our work fills this research gap by interviewing DevOps educators and allowing them to share the experience they have.

The goal of this paper is to gather challenges and recommendations reported by educators based on their DevOps teaching experience. We formulate the following research questions:

{\renewcommand\labelitemi{}
\begin{itemize}[leftmargin=*]
\item  \textbf{RQ1}: \textit{What are the challenges in DevOps courses?}
\item  \textbf{RQ2}: \textit{What are the recommendations (best practices) in DevOps courses?}
\end{itemize}
}

To answer our research questions,
we conducted an interview study with 14 DevOps educators from different parts of the world and employed thematic analysis to understand the collected data. We focus our analysis of challenges and recommendations on: (i) identification of DevOps-specific challenges and recommendations; (ii) conflicts among challenges and recommendations; and (iii) association links between them. 

Our study makes the following contributions: 

\begin{itemize}
\item A comprehensive set of 83 challenges and 185 recommendations distributed across seven themes: \textit{curriculum}, \textit{DevOps concepts}, \textit{strategies in course execution}, \textit{environment setup}, \textit{assessment}, \textit{class preparation}, and \textit{tools/technology};   
\item A comprehensive set of 558 association links between recommendations and challenges as a means to solve or mitigate challenges;
\item A set of 13 conflicting challenges and/or recommendations where more research is required to understand the adequacy of the challenges and/or recommendations to specific contexts of DevOps education.
 \end{itemize}


The remainder of this paper is structured as follows.  
Section~\ref{sec:methodology} presents our study procedure. We report and discuss challenges and recommendations in Sections~\ref{sec:results}  and~\ref{sec:discussion}.
Section~\ref{sec:threatsToValidity} reviews the threats to validity of this study. We review related work in Section~\ref{sec:relatedWork}  and we make final remarks about our work and discuss research opportunities in Section~\ref{sec:conclusion}. Moreover, we share our research artifacts publicly online~\cite{researchArtifact}.

\section{Methodology} \label{sec:methodology}

The goal of this study is to identify challenges reported by educators for teaching DevOps, as well as recommendations that help solve these challenges and to improve DevOps education. A challenge is defined as a problem that makes it difficult to plan or run a DevOps course. A recommendation is a real proposition to deal with a problem, easing the learning process.

To address our goal, the selected research methodology for this work is a cross-sectional interview study since interview-based research is appropriate for gathering in-depth descriptions of experiences, observations, and assessments~\cite{john:2017}. 

The design of this study consists of two main phases. First, semi-structured interviews with DevOps educators. Then, the analysis of the collected data to identify reported challenges and recommendations.
Six researchers conducted this work: three PhD researchers, one PhD student, one MSc student, and one undergraduate student. All of them have experience with DevOps.


The following subsections detail the two phases of our methodology.

\subsection{Interviews}


A semi-structured interview is a tool that enables the creation of an environment where the participants can share their thoughts, making it possible for the researcher to get new ideas based on interviewee's answers~\cite{hove:2005}. The goal of our interviews is to collect educators' opinions about challenges and recommendations related to teaching DevOps. 

\subsubsection{Participants Selection}
The pool of interviewees is selected from DevOps courses in GitHub, Alura\footnote{MOOC platform at  https://www.alura.com.br} (an online education platform), our personal contacts, and Google. We consider the search strings "DevOps course", "continuous delivery course", and "infrastructure as code course" in GitHub and Google.  




The criteria for selecting the interviewees are: 

\begin{itemize}
    \item the educator needs to have experience in teaching DevOps courses. The course should focus on DevOps and the educator must actively participate in the course execution;
    \item the educator has recently taught the DevOps course (maximum of two years ago). Although the Coronavirus pandemic could impair some existing courses being taught, we stipulated this period because we were interested in recent DevOps teaching experiences.
\end{itemize}

After selecting a set of possible interviewees, we invited them by sending email. For those who did not answer the first email, we sent a second one a week later. From 37 candidates invited by email, 19 demonstrated initial interest in participating of the study. However, some of them had time constraints to schedule a meeting with us or they have declined to participate after the initial contact. We finished with a total of 14 participants after discarding one candidate because the absence of interaction between this educator and the students. Table~\ref{tab:participants} summarizes educators' profiles that participated in our study.

\subsubsection{Interview Process}

The interviews were conducted and recorded using \textit{Google Meet} or \textit{Microsoft Teams} as communication technology. For recording the interview, participants need to give their permission. The 14 interviews lasted on average 32 minutes (min: 17 minutes; max: 72 minutes).

The interview starts with the presentation of the research, the interviewers, and the interviewed. Then, the main question is presented to the interviewee: \textbf{What are the challenges faced and the recommendations adopted when teaching DevOps courses?} To deepen the conversation, variations of this main question are used (see Table~\ref{tab:set_of_questions}). The creation of these variants is based on Fernandes et al.'s work~\cite{fernandes:2020}, by focusing the main question on the seven themes identified in their work. Finally, at the end of the interview, the educator is asked about any additional challenge or recommendation they may want to report.

\begin{table}[ht]
\centering
\setlength{\extrarowheight}{0pt}
\addtolength{\extrarowheight}{\aboverulesep}
\addtolength{\extrarowheight}{\belowrulesep}
\setlength{\aboverulesep}{0pt}
\setlength{\belowrulesep}{0pt}
\caption{Set of Interview Questions}
\label{tab:set_of_questions}
\begin{tabular}{|c|m{6.1cm}|} 
\toprule
\rowcolor[rgb]{0.753,0.753,0.753}

\textbf{Type} & \textbf{Question} \\ \hline

General & What are the challenges faced and recommendations adopted in general when teaching DevOps courses? \\ \hline
Theme based & What are the challenges and recommendations in the \textit{environment setup} to teach DevOps? \\ \hline
Theme based & What are the challenges and recommendations in \textit{DevOps concepts }to teach DevOps? \\ \hline
Theme based & What are the challenges and recommendations in\textit{ class preparation} to teach DevOps? \\ \hline
Theme based & What are the challenges and recommendations in \textit{tools/technologies }to teach DevOps? \\ \hline
Theme based & What are the challenges and recommendations in \textit{assessment }to teach DevOps? \\ \hline
Theme based & What are the challenges and recommendations in the \textit{curriculum }to teach DevOps? \\ \hline
Theme based & What are the challenges and recommendations in\textit{ pedagogy} (not related to curriculum nor assessment) to teach DevOps? \\ \hline
General & Any other challenges and recommendations to teach DevOps? \\ \hline
\bottomrule
\end{tabular}
\end{table}

\subsubsection{Summary of the participants}

We interviewed 14 male educators, identified as P1--P14 throughout this paper. 
As shown in Table~\ref{tab:participants}, we found 6 educators on GitHub, 3 via Google, 3 through our personal contacts, and 2 on Alura. 11 participants (78.5\%) have more than 6 years of teaching experience, and 9 (64.2\%) participants have more than 6 years of industry experience. Regarding DevOps teaching experience, 
7 (50\%) participants have been teaching DevOps for more than 2 years. We also have two educators teaching in massive open online courses (MOOCs) available throughout the year.

\subsubsection{Summary of the courses}
Table~\ref{tab:courses} shows the courses' information. The 13 courses are offered in 11 universities/companies in five countries: Brazil, USA, Canada, France, and Sweden. The majority of the courses are undergraduate and graduate level, but we also have two MOOC courses.  Although the Coronavirus pandemic required a remote teaching format, 11 courses (84.6\%) usually employ an in-person format. Moreover, 10 courses (76.9\%) are paid, only 2 courses (15.3\%) are mandatory, and 11 courses (84.6\%) have run since 2019 (see~\cite{researchArtifact}).

\begin{table}
\centering
\setlength{\extrarowheight}{0pt}
\addtolength{\extrarowheight}{\aboverulesep}
\addtolength{\extrarowheight}{\belowrulesep}
\setlength{\aboverulesep}{0pt}
\setlength{\belowrulesep}{0pt}
\caption{Participants' Profile (N=14).}
\label{tab:participants}
\begin{tabular}{|c|c|c|c|c|} 
\toprule
\rowcolor[rgb]{0.753,0.753,0.753} \textbf{Id} & \begin{tabular}[c]{@{}>{\cellcolor[rgb]{0.753,0.753,0.753}}c@{}}\textbf{Experience }\\\textbf{teaching}\\\textbf{(years)}\end{tabular} & \begin{tabular}[c]{@{}>{\cellcolor[rgb]{0.753,0.753,0.753}}c@{}}\textbf{Experience}\\\textbf{teaching}\\\textbf{DevOps}\\\textbf{(years)}\\\end{tabular} & \begin{tabular}[c]{@{}>{\cellcolor[rgb]{0.753,0.753,0.753}}c@{}}\textbf{Experience}\\\textbf{in industry}\\\textbf{(years)}\end{tabular} & \textbf{Source}                                             \\ 
\hline
P1                                            & 11-15~                                                                                                                                 & 2-3                                                                                                                                                      & 6-10                                                                                                                                     & \begin{tabular}[c]{@{}c@{}}Personal \\contact\end{tabular}  \\ 
\hline
P2                                            & 2-5~                                                                                                                                   & 2-3                                                                                                                                                      & ---                                                                                                                                      & \begin{tabular}[c]{@{}c@{}}Personal\\contact\end{tabular}   \\ 
\hline
P3                                            & 11-15~                                                                                                                                 & 1-2                                                                                                                                                      & 6-10~                                                                                                                                    & GitHub                                                      \\ 
\hline
P4                                            & 6-10~                                                                                                                                  & 2-3                                                                                                                                                      & 11-15                                                                                                                                    & Alura                                                       \\ 
\hline
P5                                            & 2-5~                                                                                                                                   & 1-2                                                                                                                                                      & ---                                                                                                                                      & GitHub                                                      \\ 
\hline
P6                                            & 6-10~                                                                                                                                  & 1-2                                                                                                                                                      & 2-5                                                                                                                                      & Alura                                                       \\ 
\hline
P7                                            & 16-20~                                                                                                                                 & 1-2                                                                                                                                                      & 16-20                                                                                                                                    & \begin{tabular}[c]{@{}c@{}}Personal\\contact\end{tabular}   \\ 
\hline
P8                                            & 6-10~                                                                                                                                  & 2-3                                                                                                                                                      & +30                                                                                                                                      & GitHub                                                      \\ 
\hline
P9                                            & 26-30~                                                                                                                                 & 2-3                                                                                                                                                      & 26-30                                                                                                                                    & Google                                                      \\ 
\hline
P10                                           & 2-5~                                                                                                                                   & 2-3                                                                                                                                                      & 21-25                                                                                                                                    & GitHub                                                      \\ 
\hline
P11                                           & 21-25~                                                                                                                                 & 1-2                                                                                                                                                      & 11-15                                                                                                                                    & Google                                                      \\ 
\hline
P12                                           & 11-15~                                                                                                                                 & 2-3                                                                                                                                                      & 16-20                                                                                                                                    & Google                                                      \\ 
\hline
P13                                           & 6-10~                                                                                                                                  & 1-2                                                                                                                                                      & 2-5                                                                                                                                      & GitHub                                                      \\ 
\hline
P14                                           & 6-10~                                                                                                                                  & 1-2                                                                                                                                                      & ---                                                                                                                                      & GitHub                                                      \\
\bottomrule
\end{tabular}
\caption*{Key: --- = Not identified.}
\end{table}

\begin{table}
\centering
\setlength{\extrarowheight}{0pt}
\addtolength{\extrarowheight}{\aboverulesep}
\addtolength{\extrarowheight}{\belowrulesep}
\setlength{\aboverulesep}{0pt}
\setlength{\belowrulesep}{0pt}
\caption{Courses' information (N=13).}
\label{tab:courses}
\begin{tabular}{|c|c|c|c|c|} 
\toprule
\rowcolor[rgb]{0.753,0.753,0.753} \textbf{Id} & \textbf{Interviewee} & \begin{tabular}[c]{@{}>{\cellcolor[rgb]{0.753,0.753,0.753}}c@{}}\textbf{Course }\\\textbf{Level}\end{tabular} & \textbf{Country} & \begin{tabular}[c]{@{}>{\cellcolor[rgb]{0.753,0.753,0.753}}c@{}}\textbf{Years of }\\\textbf{editions of }\\\textbf{the course}\end{tabular}  \\ 
\hline
C1                                            & P1                   & A                                                                                                             & Brazil           & 2016--2020                                                                                                                                   \\ 
\hline
C2                                            & P2                   & A                                                                                                             & Brazil           & 2018--2021                                                                                                                                   \\ 
\hline
C3                                            & P3                   & A                                                                                                             & Brazil           & 2018--2021                                                                                                                                   \\ 
\hline
C4                                            & P4                   & MOOC                                                                                                             & Brazil           & 2019--2021                                                                                                                                   \\ 
\hline
C5                                            & P5                   & U \& G                                                                                                          & Brazil           & \begin{tabular}[c]{@{}c@{}}2018, 2019,\\2021\end{tabular}                                                                                    \\ 
\hline
C6                                            & P6                   & MOOC                                                                                                             & Brazil           & 2020, 2021                                                                                                                                   \\ 
\hline
C7                                            & P7                   & U \& G                                                                                                          & Brazil           & 2017--2021                                                                                                                                   \\ 
\hline
C8                                            & P8                   & G                                                                                                             & USA              & 2019--2021                                                                                                                                   \\ 
\hline
C9                                            & P9                   & G                                                                                                             & USA              & 2017--2021                                                                                                                                   \\ 
\hline
C10                                           & P10, P14             & U \& G                                                                                                          & France           & 2017--2021                                                                                                                                   \\ 
\hline
C11                                           & P11                  & U                                                                                                             & Canada           & 2020, 2021                                                                                                                                   \\ 
\hline
C12                                           & P12                  & U                                                                                                             & Canada           & \begin{tabular}[c]{@{}c@{}}2019, 2020, \\2021\end{tabular}                                                                                   \\ 
\hline
C13                                           & P13                  & G                                                                                                             & Sweden           & 2019--2021                                                                                                                                   \\
\bottomrule
\end{tabular}
\caption*{Key: A = associate degree; MOOC = massive open online courses; U = undergraduate degree; G = graduate degree.}
\end{table}

\subsection{Data Extraction and Analysis}
\label{subsection:DataExtraction}

We first extracted the challenges and recommendations from interview quotes and applied the following analyses: (i) identification of DevOps-specific challenges and recommendations; (ii) thematization of recommendations and challenges; (iii) analysis of association links between challenges and  recommendations; and (iv) identification of conflicts among challenges and/or recommendations.  

Three researchers worked in collaboration to accomplish the mentioned extraction and analysis. The results of these activities are reviewed by three other researchers. During each step of the analysis, the researchers discuss and check the results before going to the next step of the analysis. In case of divergence, the researchers should avoid changing their opinion except when the first one does not find information that the second one does. If the divergence persists, an additional researcher is invited to resolve it.

\subsubsection{Transcription} 

The \textit{Telegram Transcriber Bot}\footnote{Telegram bot at   https://github.com/charslab/TranscriberBot} is used to streamline the Brazilian interviews' transcriptions. It was necessary to cut the interview file into small  pieces and process them separately because of the tool's limitations. This is not necessary  when working with \textit{Temi Transcriber}\footnote{Transcriber platform at https://www.temi.com} for the English interviews' transcriptions.

\subsubsection{Data Extraction}

We start by gathering mentions of challenges and recommendations in the interviews' transcriptions. For instance, the quote \textit{"The Docker, [...] to use, they usually have a greater difficulty in this theme, in the beginning."} (P2) is an example of a challenge where the initial difficulty is to use the Docker container tool. Similarly, the quote \textit{"We had cloud computing, where [we] can easily stand up virtual machines for people and things like that."} (P9) denotes a recommendation where cloud computing technology makes it easier to stand up virtual machines.

\subsubsection{Duplicate mentions} We consider two mentions as equal when they share the same main idea, with an essential link to join them, so they should be analyzed together. The main idea should be easily described. For example, we obtained these two quotes:
 
\begin{itemize}
    \item \textit{"We started to plan this course. And for the longest time I was really questioning myself. Like, what do you teach in a DevOps course?"} (P12)
    \item \textit{"the biggest challenge is this, like, what goes in, you know? There is a lot that people do within the DevOps Pipeline these days, which does not necessarily go into a DevOps course, right?"} (P5)
\end{itemize} 
  
In this example, both quotes refer to the challenge where there is no convention regarding the main DevOps concepts that should be taught.

\subsubsection{DevOps-specific}

We verify if DevOps concepts or the DevOps mindset appear in the challenges and the recommendations. For instance, we have the recommendation \textit{"The basics of building, testing, deploying, and monitoring should be present in a DevOps course"} (P5, P13) as an example of DevOps-specific since it is a recommendation of DevOps concepts to put into a curriculum.  On the other hand, the recommendation \textit{"Constantly try to figure out how to improve the quality of the course"} (P12) is not DevOps-specific.

\subsubsection{Themes}
We use the \textit{thematic analysis} framework proposed by Braun and Clarke~\cite{braun:2006} to create themes that summarize the data. 
{\renewcommand\labelitemi{}
\begin{itemize} [leftmargin=*]
    
    \item \textbf{\textit{Familiarization}}: The researchers start to familiarize themselves with the data in this phase. We accomplished this by reviewing the transcriptions.

    \item 

    \item \textbf{\textit{Initial Codes}}: The researchers assign codes to the data. A code is a brief description of what is being said in the interview. We separated the work into three parts for three authors, and each author followed the procedure mentioned at the start of the subsection.

    \item

    \item \textbf{\textit{Arrange Codes into Themes}}: The researchers start to sort the codes into themes. Some themes might be sub-themes of others, or some codes can become themes themselves if they are pertinent. We reused the seven themes found in Fernandes et al.~\cite{fernandes:2020}: \textit{curriculum}, \textit{DevOps concepts}, \textit{pedagogy}, \textit{environment setup}, \textit{assessment}, \textit{class preparation}, and \textit{tools}. The themes \textit{assessment} and \textit{class preparation} are sub-themes of the theme \textit{pedagogy}. In the same way, \textit{environment setup} is a sub-theme of \textit{tools}.

    \item
    
    \item \textbf{\textit{Reviewing and Defining Themes}}: The researchers reviewed and refined the initial themes. Moreover, the theme names need to be descriptive and engaging. The researchers discussed the themes and decided to make the following changes: 
	i) the theme \textit{pedagogy} changed to \textit{strategies in course execution} because we identified that the corresponding codes are related to the context during the course execution; 
	ii) the theme \textit{tool} changed to \textit{tools/technology} because we want to make the extension of this theme explicit.

\end{itemize}
}

\subsubsection{Association Links} 

We consider an association link between a challenge and a recommendation when the recommendation solves or mitigates the challenge. As an example: we can link the recommendation \textit{"Build scenarios that students can run on their computer"} (P1) to the challenge \textit{"Using cloud services more professionally requires payment at a commercial level"} (P14). Here the idea is to run applications from the course on a student's computer instead of using paid cloud services.

During the identification of association links, we first searched by themes, considering that there is a higher likelihood to find recommendations that can address challenges from the same theme.

\subsubsection{Inconsistencies or conflicts}

We find an inconsistency when there is a disagreement between a challenge and a recommendation or between two recommendations. This can happen because the challenges and recommendations reflect the educator's experience in their specific DevOps course, which might be inconsistent with other educators' experience. Similar to the identification of association links, we search first by themes.

\section{Results}  \label{sec:results}




The average duration of the interviews was 32.9 minutes (standard deviation$=$14.6).
We provide a set of artifacts~\cite{researchArtifact} containing all data collected and analyzed in this paper.

\begin{figure}[ht]
  \includegraphics[width=8.5cm]{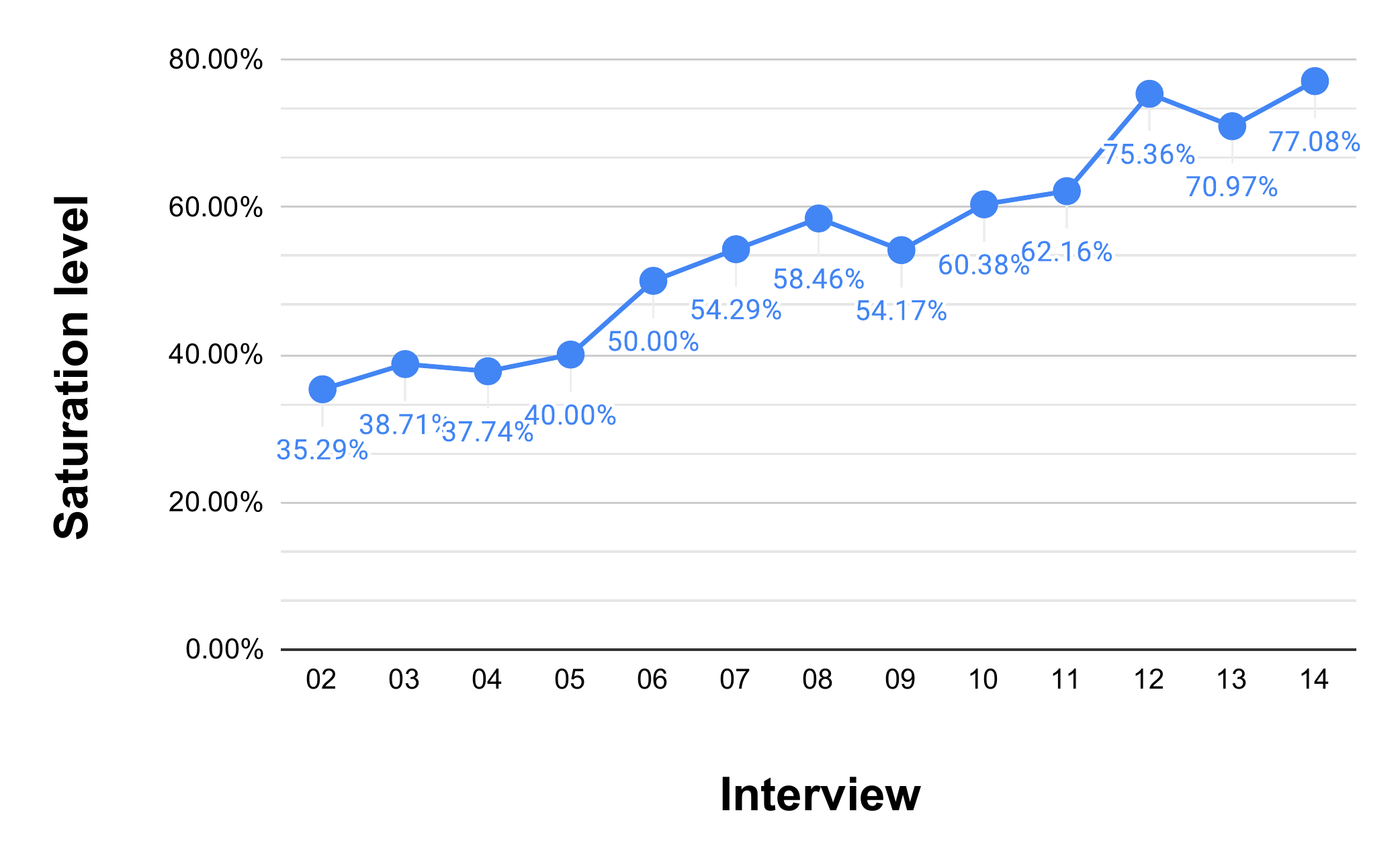}
  \caption{For each interview, percentage of challenges and recommendations already mentioned in the previous interviews.}
  \label{figura:duplicationChallengesRecommendations}
\end{figure}

Figure~\ref{figura:duplicationChallengesRecommendations} shows the saturation level of challenges and recommendations throughout the interview process. The saturation of an interview is the percentage of challenges and recommendations that had
already been identified in previous interviews.
We concluded the interview phase when new interviews only led to minor changes to themes. As Figure 1 shows, this coincided with a saturation level of about 80\%.

\subsection{Overview of Challenges and Recommendations}

We identified 83 challenges and 185 recommendations for DevOps teaching. Due to space limitation, we present an overview of the main challenges and recommendations found. The complete results are available in our online appendix~\cite{researchArtifact}.


Table~\ref{tab:challengesMostCited} shows the ten most mentioned challenges.  About 12\% of the challenges were mentioned by three or more interviewees. Eight educators mentioned the \textit{"Insufficient time in the course to teach DevOps"} (P1, P3, P5, P6, P8, P9, P10, P11) challenge. Table~\ref{tab:challengesMostCited} also illustrates that some of the mentioned challenges are related to each other and were grouped in the table. For example, the challenges \textit{"It's hard to show to students that DevOps is not all about tooling"} (P3, P4, P6, P9, P11, P12, P14), \textit{"There is a large number of DevOps tools"} (P2, P3, P5, P6, P7, P8, P11), and \textit{"It's challenging to be up-to-date with industrial DevOps tools"} (P5, P7, P8, P9, P10, P13, P14), are all related to the usage of a large set of tools in DevOps.

\begin{table}[ht]
\centering
\setlength{\extrarowheight}{0pt}
\addtolength{\extrarowheight}{\aboverulesep}
\addtolength{\extrarowheight}{\belowrulesep}
\setlength{\aboverulesep}{0pt}
\setlength{\belowrulesep}{0pt}
\caption{Most Mentioned Challenges}
\label{tab:challengesMostCited} 
\begin{tabular}{|m{6.1cm}|c|}
\toprule
\rowcolor[rgb]{0.753,0.753,0.753}
\multicolumn{1}{|c|}{\textbf{Challenge}} & \begin{tabular}[c]{@{}>{\cellcolor[rgb]{0.753,0.753,0.753}}c@{}}\textbf{Number of }\\\textbf{occurrences}\end {tabular}  \\ \hline
Insufficient time in the course to teach DevOps. (P1, P3, P5, P6, P9, P10, P11) & 8 \\ \hline
There is a large number of DevOps tools. (P2, P3, P5, P6, P7, P8, P11) & 7 \\ \hline
It's hard to show to students that DevOps is not all about tooling. (P3, P4, P6, P9, P11, P12, P14) & 7 \\ \hline
It's challenging to be up-to-date with industrial DevOps tools. (P5, P7, P8, P9, P10, P13, P14) & 7 \\ \hline
It's challenging to balance DevOps theory and practice. (P2, P4, P6, P11, P12, P14) & 6 \\ \hline
Insufficient knowledge level of students to start the course. (P1, P4, P5, P8, P9) & 5 \\ \hline
It's challenging to deal with students having different backgrounds. (P4, P5, P6, P8, P10) & 5 \\ \hline
DevOps culture is hard to teach. (P3, P4, P6, P8) & 4 \\ \hline
Teaching DevOps concepts to students with no industrial experience is hard. (P3, P12, P13, P14) & 4 \\ \hline
It's challenging to find the right sized examples to teach DevOps. (P5, P9, P11, P14) & 4 \\ \hline
\bottomrule
\end{tabular}
\end{table}

Table~\ref{tab:recommendationMostCited} shows the ten most mentioned recommendations.  About 10\% of them were mentioned by more than three educators. Eight educators recommended a \textit{"teaching method based on practical activities"} (P1, P2, P3, P5, P6, P7, P10). Comparing Tables~\ref{tab:challengesMostCited} and~\ref{tab:recommendationMostCited}, we can see the importance of teaching the DevOps mindset although it is a hard task. We also note the recommendation \textit{"Use relevant industry tools"} (P6, P7, P8, P9, P11, P12) as a strategy to address the challenge \textit{"There is a large number of DevOps tools"} (P2, P3, P5, P6, P7, P8, P11). Moreover, four interviewees recommended \textit{"Teach using examples"} (P3, P4, P11, P12), but it is not easy to find the right sized ones.

\begin{table}[ht]
\centering
\setlength{\extrarowheight}{0pt}
\addtolength{\extrarowheight}{\aboverulesep}
\addtolength{\extrarowheight}{\belowrulesep}
\setlength{\aboverulesep}{0pt}
\setlength{\belowrulesep}{0pt}
\caption{Most Mentioned Recommendations}
\label{tab:recommendationMostCited}
\begin{tabular}{|m{6.2cm}|c|} 
\toprule
\rowcolor[rgb]{0.753,0.753,0.753}
\multicolumn{1}{|c|}{\textbf{Recommendation}} & \begin{tabular}[c]{@{}>{\cellcolor[rgb]{0.753,0.753,0.753}}c@{}}\textbf{Number of }\\\textbf{ocurrences}\end{tabular}  \\ \hline
Teaching method based on practical activities. (P1, P2, P3, P5, P6, P7, P10) & 7 \\ \hline
Use cloud provider services. (P4, P7, P8, P9, P12, P14) & 6 \\ \hline
Use relevant industry tools. (P6, P7, P8, P9, P11, P12) & 6 \\ \hline
Use cloud provider services with students plans. (P1, P3, P4, P7, P12) & 5 \\ \hline
The assessment should be with hands-on activity. (P1 P2, P3, P5, P7) & 5 \\ \hline
Use Jenkins tool. (P4, P5, P10, P11, P14) & 5 \\ \hline
Teach continuous integration and pipeline automation. (P5, P8, P12, P13, P14) & 5 \\ \hline
Teach the DevOps mindset. (P3, P4, P8, P12) & 4 \\ \hline
Focus more on the practical part compared to the theoretical part of DevOps. (P1, P8, P10, P11) & 4 \\ \hline
Teach using examples. (P3, P4, P11, P12) & 4 \\ \hline
\bottomrule
\end{tabular}
\end{table}

\begin{figure}[ht]
  \includegraphics[width=8.5cm]{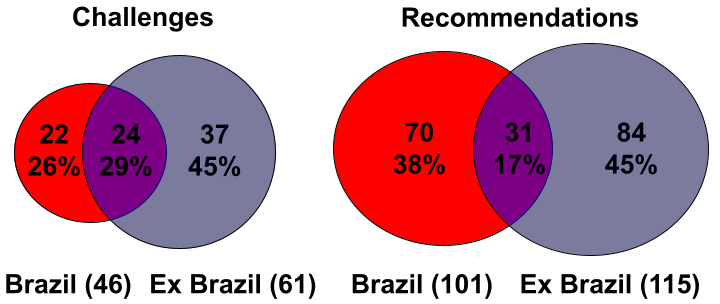}
  \caption{Distribution of Challenges and Recommendations mentioned by Interviewees Inside and Outside Brazil}
  \label{figura:brazilDevOps}
\end{figure}

Half of the 14 interviewees are from Brazil. As shown in Figure~\ref{figura:brazilDevOps}, the minority of challenges (29\%) and recommendations (17\%) are present in both demographics inside Brazil and outside Brazil. Such as, for example, the challenge \textit{"Insufficient knowledge level of students to start the course."} (P1,P4,P5,P8,P9) and the recommendation \textit{"Use cloud provider services with students plans."} (P1,P3,P4,P7,P12). However, only Brazilian interviewees mentioned the challenge \textit{"Unknown unified material for teaching DevOps."} (P5,P7) and the recommendation \textit{"Use various sources of DevOps materials."} (P5,P7). Moreover, only interviewees outside Brazil reveal the challenge \textit{"The preparation of the exercise is demanding."} (P10,P11) and the recommendation \textit{"Select industrial speakers carefully to share their experience with the students."}. (P12,P13,P14)

\subsection{DevOps-specific Challenges and Recommendations}

We realized that some of the challenges and recommendations mentioned by interviewees arise in situations to teaching DevOps concepts and culture while others are about education in general.

\begin{figure}[ht]
  \includegraphics[width=.49\columnwidth]{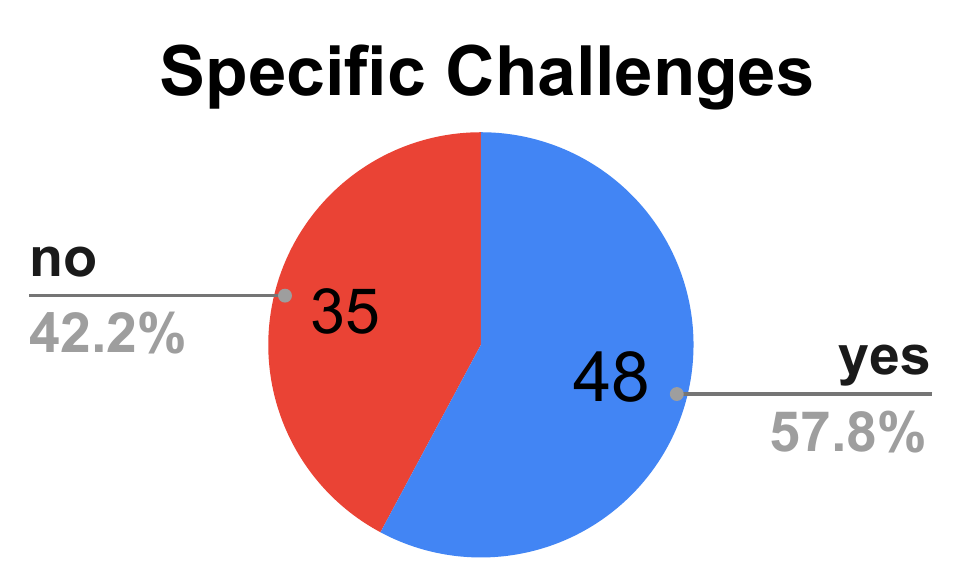} \hfill
  \includegraphics[width=.49\columnwidth, height=2.55cm]{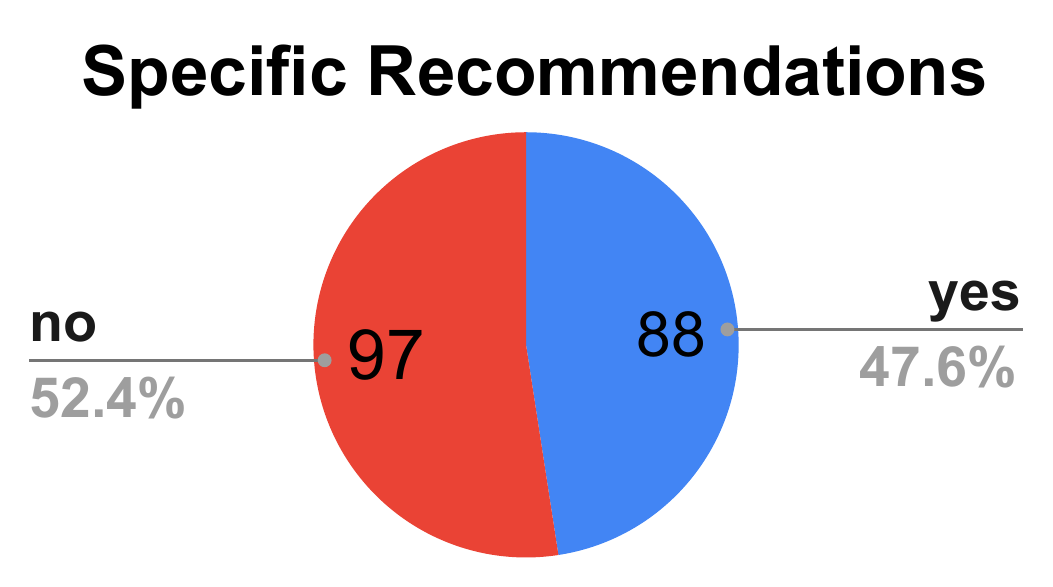}
  \caption{Distribution of DevOps-specific Challenges and Recommendations}
  \label{figura:specificDevOps}
\end{figure}

Figure~\ref{figura:specificDevOps} shows the distribution of DevOps-specific challenges and recommendations. The majority (57.8\%) of the challenges are related to DevOps, while 47.6\% of the recommendations are DevOps-specific. For instance, the \textit{"Difficulty in assessing students' understanding of Continuous Delivery"} (P2) challenge is DevOps-specific because continuous delivery is a DevOps concept. However, \textit{"Students heavily rely on the professor's slide material, which is often limited"} (P5) is not considered DevOps-specific because it is related to students' behaviour in general.

In the same way, the recommendation \textit{"Use Jenkins, Travis CI, Circle CI and GitHub Actions to teach Continuous Integration"} (P5) is identified as DevOps-specific, because continuous integration is one of the DevOps practices. Meanwhile, \textit{"It is necessary to choose which topics and tools are essential as the course time is limited"} (P6) is not DevOps-specific because the selection of course topics is a common educational concern.

\subsection{Thematization of Challenges and Recommendations}

\begin{figure}[ht]
  \includegraphics[width=8.5cm]{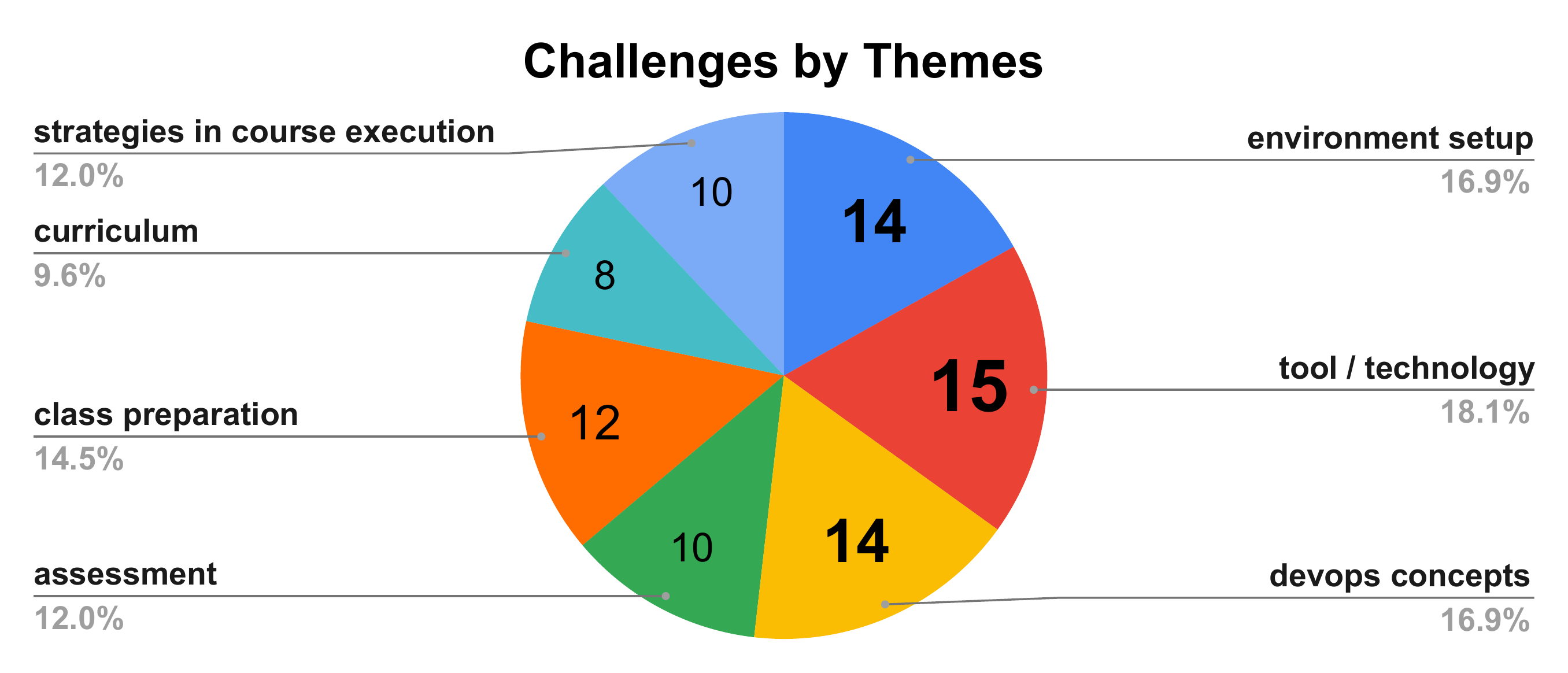}
  \caption{83 challenges spread over 7 themes}
  \label{figura:DesafiosTemas}
\end{figure}
\begin{figure}[ht]
  \includegraphics[width=8.5cm]{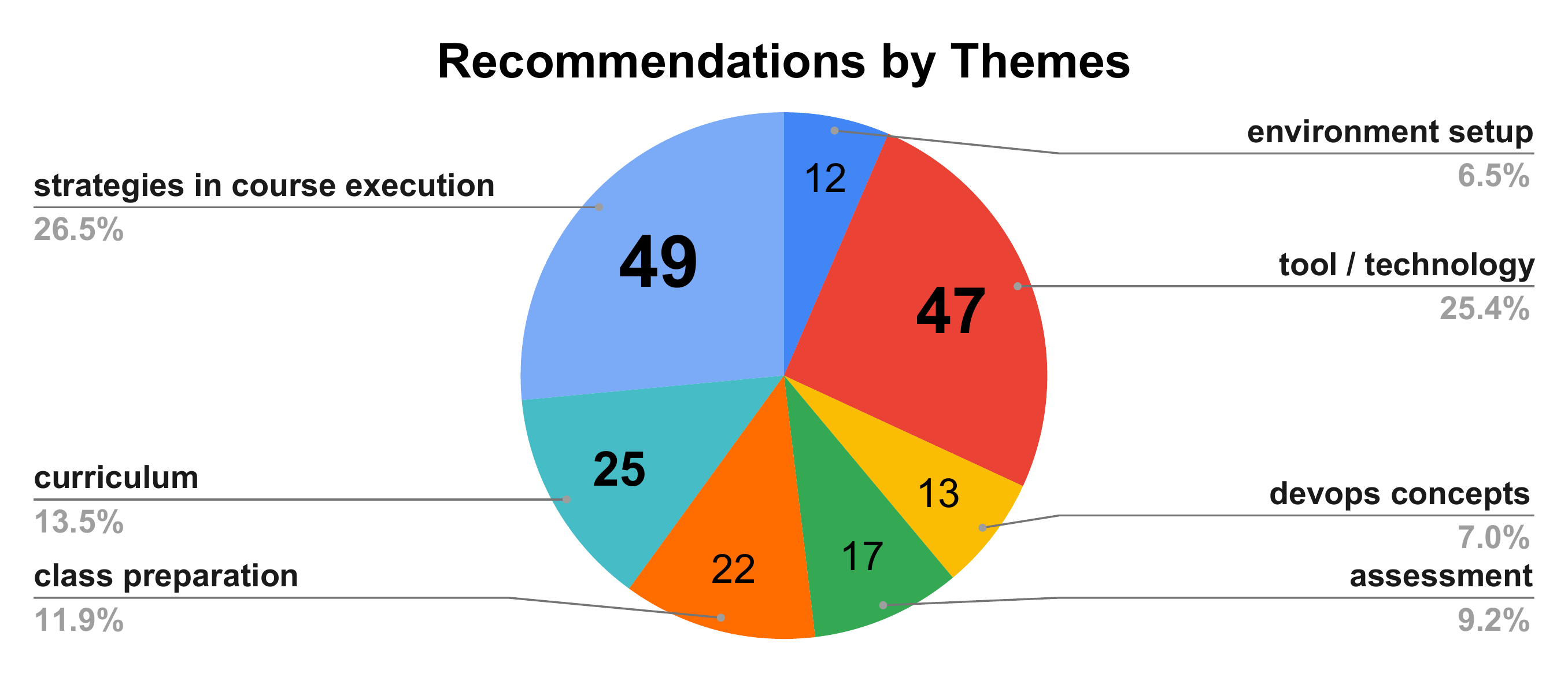}
  \caption{185 recommendations spread over 7 themes}
  \label{figura:RecomendacoesTemas}
\end{figure}

Our thematic analysis resulted in seven themes which cover both challenges and recommendations. The themes are: \textit{Strategies in Course Execution}, \textit{Curriculum}, \textit{Assessment}, \textit{Tool/Technology}, \textit{Class Preparation}, \textit{DevOps concepts}, and \textit{Environment Setup}. Figure~\ref{figura:DesafiosTemas} and Figure~\ref{figura:RecomendacoesTemas} illustrate the distribution of these themes across the challenges and recommendations, respectively. 

Figure~\ref{figura:DesafiosTemas} shows a balance between the total number of challenges for the following themes: \textit{Environment Setup}, \textit{Tool/Technology}, and \textit{DevOps concepts}. The number of challenges for these themes together represents about 50\% of the total. We also identified \textit{Tools/ Technology} and \textit{Curriculum} as the most and least recurrent themes, respectively. 

On the other hand, Figure~\ref{figura:RecomendacoesTemas} shows that there is no balance between the number of recommendations under each theme. For example, the \textit{Strategies in Course Execution} theme covers about 26\% of the recommendations. \textit{Strategies in Course Execution} and \textit{Tool/Technology} represent about 51\% of the total of recommendations. 

We can observe from the results presented in Figures~\ref{figura:DesafiosTemas} and \ref{figura:RecomendacoesTemas} that \textit{Tool/Technology} is a recurrent theme for educators. Moreover, we noticed that \textit{Environment Setup} is a recurrent challenge, but only a few recommendations were provided by the interviewed educators to address this challenge.

The following subsections present results of the analysis grouped by theme. In particular, we show how the identified challenges and recommendations were derived from specific situations mentioned during the interviews.  


\subsubsection{Strategies in Course Execution}

This theme is related to the set of techniques, principles, methods, and procedures of education and teaching that are used during the course execution.

\vspace{2mm}

\textit{\textbf{Challenges.}} P5 explained \textit{"You have people who are very proficient in development and have no idea about server, linux and environment configuration. People who came from operational and are not so proficient in the programming part"}, and P6 added \textit{"Each student brings a different experience, different challenges and trying to generalize this is more complicated."}. These quotes are related to the same challenge of \textbf{dealing with students with different backgrounds}.

P5 described \textit{"People studied a lot from the material I prepared. When the material I was preparing was, let's say, it was a set of slides."} and P12 added \textit{"I actually strongly encourage students to read the book, but I know that they don't."}. These quotes represent the challenge that \textbf{students rely on limited material}.



\vspace{2mm}

\textit{\textbf{Recommendations.}} \textbf{Teaching method based on practical activities} is the recommendation obtained from quotes such as: \textit{"So, didactically, pedagogically, teaching in this way, with the most practical approach, I think it's better."} (P2), \textit{"I can't see a discipline, a DevOps teaching that isn't hands-on"} (P5).

\textbf{Show use cases of DevOps} is the recommendation illustrated in the following quotes: \textit{"Show use case about elimination of silos among the development and operations team.} (P3), and \textit{"The lectures are more the, you know, sharing the experience."} (P12).



\subsubsection{Curriculum} 

This theme relates to the content, workload, and number of courses required to use DevOps. The theme considers the relationship of the DevOps discipline with prerequisite disciplines, as well as the interaction with the rest of a software engineering program.

\vspace{2mm}

\textit{\textbf{ Challenges.}} P1 explained \textit{"A recurrent problem is the level of students' knowledge that they [have] when they start the discipline."}, and P5 added \textit{"The lack of proficiency of some students, in some criteria of this, ends up making this practice difficult."}.
These quotes are related to the challenge of \textbf{Insufficient knowledge level of students to start the course}.

P1 told us \textit{"We sometimes want to teach everything and we don't have infinite time"}, and P6 said \textit{"So training is limited. It's a forty-hour training, right?"}.
These quotes represent the challenge \textbf{Insufficient time in the course to teach DevOps}.


\vspace{2mm}

\textit{\textbf{ Recommendations.}} \textbf{The courses of software architecture and DevOps taught together} is the recommendation obtained from quotes such as \textit{"So we built a curriculum in just very innovative way, the DevOps and Software Architecture classes together, a single project, a single teaching team, but we evaluate on two angles."} (P10) and \textit{"They were really like Friday was dedicated to DevOps and architecture."} (P14).

\textbf{Prepare students with previous courses that teach related DevOps concepts} is the recommendation illustrated in the following quotes: \textit{"It'd be great if there was a cloud course before mine, but there isn't."} (P8) and \textit{"It's an option that we give them the year before to preparing them."} (P10).


\subsubsection{Assessment}

This theme includes mandatory and permanent tasks in the professor's work to assess each student's learning situation concerning the curriculum schedule.

\vspace{2mm}

\textit{\textbf{Challenges.}} P3 said \textit{"In many cases the assessment is still based on the traditional test model."} to warn about the \textbf{problem of dealing with assessments based on a traditional test model}.

P10 described \textit{"I check out the code of every group. And I look at the commits, who has done what, I look at. How has it been coded, copy paste of somebody else's code? Is it innovative? I run all the scripts. ... And I run them on my computer. It's very long."}. He illustrates \textbf{how arduous it is to analyze the code of each project}.



\vspace{2mm}

\textit{\textbf{ Recommendations.}}
\textbf{Grade students based on their learning journey and mistakes} is the recommendation obtained from quotes such as, \textit{"It would help if you observed each student and, and then, you have to listen a lot, too, what was the difficulty he had and where he arrived."} (P4), and \textit{"I teach them that every failure is a learning opportunity. [...] we're here to learn."} (P8).

\textbf{The assessment should be with a hands-on activity} is the recommendation illustrated in the following quotes: \textit{"You can't evaluate with test; you have to assess with projects with some activity."} (P1), and \textit{"the evaluation of this project puts a student in his context to test in practice or simulate, in practice, a little of what he saw during classes."} (P3).




\subsubsection{Tool/Technology}
This theme includes the tools and technologies used to operationalize DevOps practices. This includes tools and technologies used widely in industry or those created for teaching purposes.

\vspace{2mm}

\textit{\textbf{Challenges.}} P3 noted \textit{"In general you have a wide range of solutions, you have a very large ecosystem of possibilities on how to test or demonstrate a concept."}, P5 explained \textit{"A difficulty with technologies is recognizing what is relevant to be addressed in the classroom, isn't it? So, for example, there is a lot of technology on the market."}, and P6 elaborated \textit{"Because the DevOps universe has millions of tools, technologies and such."}.
These quotes represent the challenge \textbf{There is a large number of DevOps tools.}.

P14 told us \textit{"The lab session, they have to be really precise. You [..] work one day. And then the second day it does not work because there is an upgrade in the Docker API that makes things totally different"}. In the same way, P8 explained, \textit{"So the challenge for me is that the cloud is constantly evolving. And so every semester what I try to do in my class, in my labs, I have snapshots of screenshots and circles and arrows and, you know, click on this and move there. And that changes constantly"}.  Both quotes illustrate the challenge of \textbf{being up-to-date with industrial DevOps tools}.



\vspace{2mm}

\textit{\textbf{ Recommendations.}} 
\textbf{Use cloud provider services with student plans}. In this context, P4 said, \textit{"I recommend [...] moving all teaching to a cloud. [...] contact AWS, they have a student program, or Google, with Ali Baba, with Azure, with the IBM Cloud"}. Corroborating this, P1 shares, \textit{"Amazon has some agreements [...] that provide student accounts that they can test for a period"}.


\subsubsection{Class Preparation}
This theme refers to course planning, including, for example, searching for reference material and preparing lessons.

\vspace{2mm}

\textit{\textbf{Challenges.}} 
P7 shared \textit{"The teaching plan [...] how are you going to connect subjects, [...] is the hardest part"}. This interviewee added, \textit{"What they are going to do, was the very difficult part [...] So, you cannot think about doing a theoretical thing, you have to have practice, you cannot just to be just practical exercises, it has to have a whole journey, a well-established train of thought. It was quite tricky to get to that topic"}. These quotes refer to the challenge of \textbf{difficulty to structure the learning journey}.




\vspace{2mm}

\textit{\textbf{Recommendations.}}
P4 explained, \textit{"Suppose it is a class that, specifically, we were given the needs and characteristics before, such as access limitations, limited software installation on the machine. In that case, I prepare the class, and we have the schedule as a whole"}. This quote refers to the recommendation to \textbf{seek to know in advance the needs and limitations of the class}.

P7 explained that \textit{"most of the references, the most interesting cases that I considered to bring to the room are posts on INFO2, on Metzone, Hacker News, Twitter posts, Airbnb case study, Glitch, Orbitz and such; other cases of those that are much more interesting than necessarily, books or 'scientific academic' articles"}. We formulate the corresponding recommendation as \textbf{use various sources of DevOps materials}.


\subsubsection{DevOps Concepts} 
This theme is related to fundamentals, techniques, and mindset (culture) of DevOps.

\vspace{2mm}

\textit{\textbf{Challenges.}}
P9 said, \textit{"People coming through the programs want to play with technology [...] But what that tends to foster is a technology-centric attitude about what DevOps is all about"}. Similarly, P3 said, \textit{"The student hopes to [...] learn that killer tool, which will help in the practical context of his life, whether in the process of development, security, or operations. [...] wants to know the tools much more than understand the DevOps culture."}. 
These quotes illustrate the challenge of the \textbf{difficulty in showing students that DevOps is not all about tooling}.

P9 shared \textit{"The challenge, of course, is newer students obviously have more than enough to worry about just getting code wrong and compiling. But that is the reality, unfortunately, is the code does not run a compile on a laptop, right? It runs out in production, and it is serving real people. And in this day and age, there is stuff that goes with that. And the more folks understand, at least some of the sooner, the better I hope the software will be"}. This quote refers to the \textbf{difficulty for students to understand the importance of software running in production, not just compiling}.



\vspace{2mm}

\textit{\textbf{Recommendations.}}
P8 recommended, \textit{"working as an agile team and using the DevOps tools, but most importantly, living the DevOps culture."} and \textit{"The only way to teach culture, the only way to experience the culture is to immerse the students in the culture. [...] We are going to experience DevOps. And that is the only way you can properly teach it"}. In the same way, P3 added, \textit{"Always focus on culture, the tools are excellent, they attract a student, they create a practical scenario, but oh, DevOps implementation errors in practice are mainly caused by companies and professionals who do not interpret this as a culture"}.   These quotes emphasize the importance of \textbf{teaching the DevOps mindset}.



\subsubsection{Environment Setup}

This theme refers to the preparatory activities related to the environments used in exercises and student projects.

\vspace{2mm}

\textit{\textbf{ Challenges.}}
P1 told us, \textit{"I had difficulty setting up the infrastructure."}, and P14 added, \textit{"We have tried to let the students deal with the setup and install everything on their computer with Dockerizing stuff and scan things. And that was yet another disaster because then it is not reproducible and it works on their computer, but then it is really complicated to make it work on the [computer of the] teaching assistant"}.
These quotes describe the challenge where \textbf{infrastructure setup is difficult}.





\vspace{2mm}

\textit{\textbf{  Recommendations.}}
P1 recommended, \textit{"Set up scenarios that they can run on their computer."}, \textit{"adapt to something perhaps less computationally demanding."}, \textit{"sometimes give up certain things you would like to teach [...] to the detriment that the student does not have the ability to perform"}. These quotes are about the same recommendation of \textbf{building scenarios that students can run on their computers}.	

P5 added, \textit{"Maybe it makes sense for you to provide the environment for the students, right? And this provision, you can use a Docker related technology, which comes already, right?"}. This quote refers to \textbf{providing an initial environment setup for students}.	



\subsection{Linking DevOps Recommendations to Challenges}

In addition to analyzing and extracting the challenges and recommendations from the interviews conducted in our study, we also identified association links between DevOps challenges and recommendations. The links were obtained as a direct result of the suggestions of the educators during the interviews, but we also establish links between challenges and recommendations through several rounds of discussion. 

\begin{center}
\begin{figure*}[h]
  \includegraphics[width=17cm]{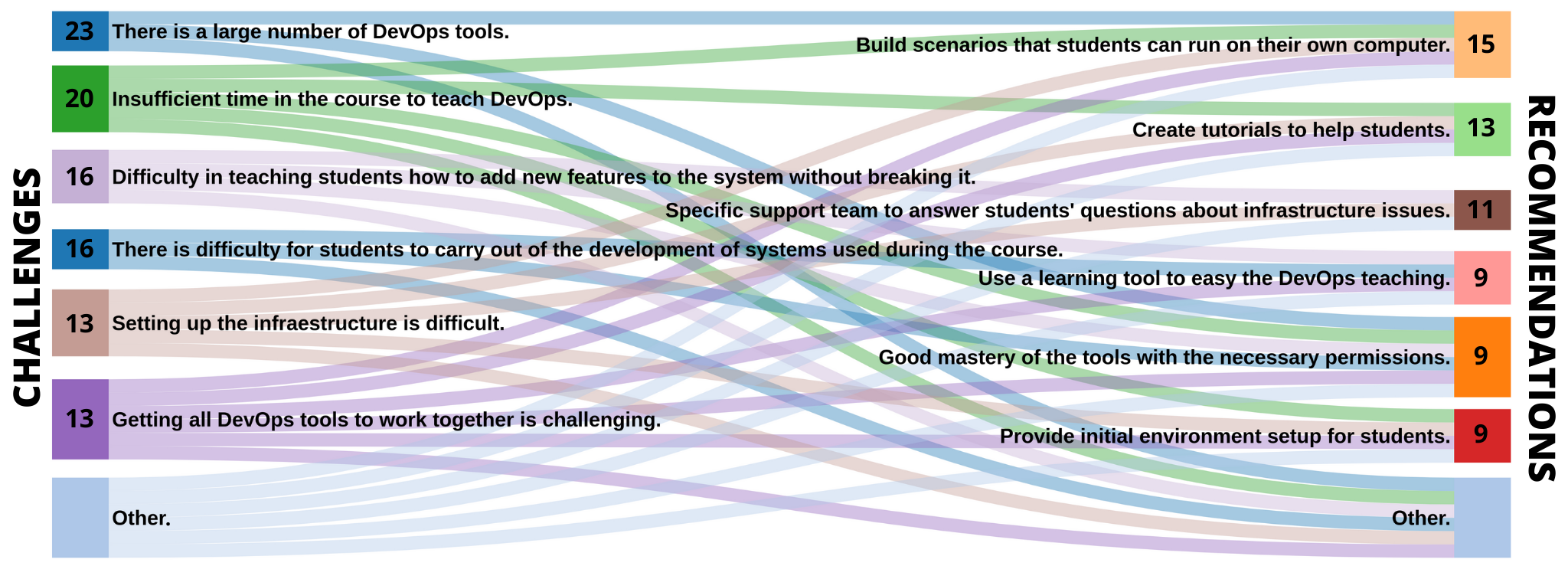}
  \caption{Most challenges and recommendations linked.}
  \label{figura:sankeyDiagram}
\end{figure*}
\end{center}

This analysis resulted in a total of 558 links between challenges and recommendations. Figure~\ref{figura:sankeyDiagram} presents the challenges and recommendations most linked through a Sankey diagram. The challenge \textit{"There is a large number of DevOps tools"} (P2, P3, P5, P6, P7, P8, P11) has 23 association links with recommendations, i.e., 23 recommendations help to mitigate or solve the challenge. This challenge is related to, for example, the following recommendations: \textit{"Build scenarios that students can run on their own computer."} (P1), \textit{"When teaching a tool, you must understand their main commands and have the necessary permissions to deal with it"} (P2), and \textit{"Use relevant industry tools."} (P6, P7, P8, P9, P11, P12). Another example is the challenge \textit{"Insufficient time in the course to teach DevOps."} (P1, P3, P5, P6, P8, P9, P10, P11) that has 20 links to recommendations, such as: \textit{"A discipline must have a considerable workload to centralize and harmonize development and operation information."} (P2), \textit{"Know in advance the needs and limitations of the students."} (P4, P5), and \textit{"It is necessary to choose which topics and tools are essential as the course time is limited."} (P6).

Considering all challenges, the median value is eleven challenges linked to four recommendations. Only the challenge \textit{"It is really difficult to quantitatively grade scale on the description and the justification of case studies"} (P14) has no link. 


From another perspective, the recommendation \textit{"Build scenarios that students can run on their own computer."} (P1) is associated with 15 challenges, i.e., it mitigates or solves 15 challenges, such as \textit{"Cloud providers usage has limits."} (P1, P5, P14) and \textit{"Debugging lab sessions are very difficult."} (P14). Another example is the recommendation \textit{"Try to make the environment setup minimal."} (P12), which has 13 links with associated challenges such as \textit{"Setting up the infrastructure is difficult."} (P1, P7, P14) and \textit{"Institutions' resources have limits"} (P1, P9, P10).

Considering all recommendations, the median value is 39 recommendations linked to two challenges. We also find a total of 28 recommendations that were not associated to any challenge.


\subsection{Conflicts between Challenges and/or Recommendations}  \label{subsection:conflicts}

During our analysis, we also identified scenarios that represent conflicts between challenges and/or recommendations, thus introducing inconsistency to their application. Below, we present and discuss these conflicts.

\begin{enumerate}[label=\roman*]
    \item The recommendation \textit{"Organize the curriculum equally between concepts and tools"} (P3) prescribes a different curriculum division than the recommendation \textit{"Focus more on the practical part than DevOps' theoretical aspects"}.
    \item One challenge \textit{"Teaching many tools makes the student think that DevOps is just about tools"} (P9) warns about using too many tools. The recommendation \textit{"Use as many as possible, prioritizing those used in industry for future use at work"} (P5) is in conflict with it.
    \item Two interviewees recommends \textit{"Restricting the use of the discipline's technology stack"} (P10, P11), but four interviewees suggest no restriction (P8,P12,P13,P14).
    \item The challenge \textit{"It is difficult to deal with assessments based on a traditional model"} (P3) warns about using a traditional assessment method. However the recommendation \textit{"Do exams with more conceptual questions"} (P5, P8, P9) suggests to use a traditional assessment method.
    \item Eight interviewees have the challenge \textit{"The course time is insufficient to teach DevOps"} (P1, P3, P5, P6, P8, P9, P10, P11), but one says the time is enough (P13).
    \item Although we identified the challenge \textit{"No time to practice on Kubernetes"} (P10), one interviewee recommends \textit{"Only 20\% of the course to teach applications"} (P11).
    \item One recommendation says \textit{"The assessment should be with a hands-on activity"} (P1, P2, P3, P5, P7). However, another proposes \textit{"To do exams with more conceptual questions"} (P5, P8, P9).
    \item The recommendation \textit{"Use simple tools"} (P6, P10) is in conflict with the recommendation about using Jenkins (P4, P5, P10, P11, P14), a not-so-simple tool.
    \item One recommendation says \textit{"Small examples were not really satisfactory"} (P11) to teach DevOps. Another suggests \textit{"Use small examples"} (P9, P10, P12).
    \item We identified the challenge \textit{"The process of making students migrate to other tools is hard"} (P2). However, we also obtained the recommendation \textit{"Ask students to adopt the tools used by instructors"} (P2).
    \item We cannot define the ideal number of tools to teach DevOps, given the conflicting recommendations \textit{"Use many tools"} (P5) and \textit{"Use only the most relevant ones"} (P6, P7, P8, P9, P11, P12).
    \item The challenge \textit{"There is insufficient literature related to teaching DevOps"} (P2, P5, P11) is in conflict with the recommendation \textit{"Use a textbook as a basis to guide the course classes"} (P2, P12).
    \item Although we identified the challenge \textit{"It is difficult to deal with assessments based on a traditional test model"} (P3), one recommendation is \textit{"To use a quiz with multiple choices to assess the students"} (P8, P12).
\end{enumerate}
All of these conflicts present opportunities for future work, focusing specifically on the aspects of DevOps education that our interviewees did not agree on.

\section{Discussion}  \label{sec:discussion}

We discuss our results by comparing them with findings previously obtained from a systematic review and in terms of the results' implications for educators and researchers.

\subsection{Comparison with Systematic Review}

Compared to the systematic review of 18 primary papers presented by Fernandes et al.~\cite{fernandes:2020}, our work identified substantially more information based on 14 interviews. A quantitative comparison (this work vs.~Fernandes et al.'s work~\cite{fernandes:2020}) is: 83 vs.~73 challenges, 185 vs.~85 recommendations, 558 vs.~149 association links between challenges and recommendations, and 13 vs.~2 conflicts among challenges and recommendations. Moreover, this study highlights more DevOps-specific challenges (48 vs.~26) and DevOps-specific recommendations (88 vs.~23). 60\% (44 of 73) of the challenges and 54\% (46 of 85) of the recommendations from Fernandes et al.'s systematic review appear in our work. In turn, our interview-based study identified 39 new challenges and 139 new recommendations.

Similar to Fernandes et al.~\cite{fernandes:2020}, we identified several challenges related to the preparation and usage of DevOps tools to support teaching activities, such as \textit{"There is a large number of DevOps Tools"} (P2, P3, P5, P6, P7, P8, P11), \textit{"It’s hard to show to students that DevOps is not all about tooling"} (P3, P4, P6, P9, P11, P12, P14), and \textit{"It’s challenging to be up-to-date with industrial DevOps tools"} (P5, P7, P8, P9, P10, P13, P14). However, our study extends the number of recommendations related to those challenges. A total of 22, 10, and 3 new recommendations were found to address these challenges, respectively, e.g., \textit{"Use relevant industry tools"} (P6, P7, P8, P9, P11, P12), \textit{"Start teaching DevOps from the culture. Then demonstrate with tools"} (P3), and \textit{"Someone from professor staff implements the sample application."} (P11). 

The most recurrent recommendation in Fernandes et al.~\cite{fernandes:2020} is \textit{"Teaching method based on practical activities"} (P1, P2, P3, P5, P6, P7, P10). This recommendation is also the most recurrent one in our work. Half of the educators (7 of 14) mentioned the importance and adoption of practical activities to exercise concepts such as Continuous Delivery. Some of them also mentioned \textit{"the need to define such practical exercises based on constraints of the scenario"} (P1, P2, P3, P5, P7).

The second most mentioned recommendation by Fernandes et al.~\cite{fernandes:2020} is \textit{"The employing of non-traditional forms of teaching such as peer instruction, inverted classroom instruction, and problem-based learning"} (P7). Our study also found recommendations such as \textit{"Problem-based learning (PBL) is adequate for teaching DevOps, and it could be merged with inverted classes and agile"} (P7).

As the results of the two studies are comparable, we recommend further investigation of the challenges and recommendations in DevOps Education, including themes, links, and conflicts.

Pang et al.~\cite{pang:2020} discussed the challenge \textit{"To configure such infrastructure for many students in an academic environment"} (P1, P7, P14). Our study reinforced this aspect with many challenges about the theme \textit{environment setup} such as \textit{"Difficulty in understanding environment, tools and network configuration"} (P5), \textit{"Prepare the labs environment requires a lot of time."} (P5, P12, P14), and \textit{"You need a lot of interconnected machines running different services with visibility on each other to do continuous deployment."} (P10). We also found many recommendations linked to these challenges, such as \textit{"Create tutorials to help students"} (P4, P10, P12), \textit{"Try to make the environment setup minimal"} (P12), and \textit{"Build scenarios that students can run on their own computer"} (P1).



\subsection{Implications to Educators} 

Our study has identified a large set of DevOps challenges and recommendations which we have analyzed and organized by themes, links, and conflicts that can help educators with planning, execution, and evaluation of DevOps courses. This in itself can be useful material to mitigate the challenge \textit{"Insufficient literature focused on teaching DevOps"} (P2, P5, P11).

In our study, we also observe the need for educators to prepare their educational methodologies to motivate students to understand the DevOps mindset and not only learn technical/technological aspects. For example, one of our interviewed professors, who also works in industry, mentioned the following recommendation: \textit{"I tell them, I am not going to grade you on what you submit. I'm going to grade you on how you got there because getting there is not the point. It's the journey, right? That's the point. [...] I teach them that every failure is a learning opportunity. If you fail and you learn something, you get credit. It's not a failure because you've learned something, we're here to learn."} (P11) 

Educators can use the recommendations linked with challenges to avoid and mitigate a common set of problems found when teaching DevOps. The recommendations provide a set of best practices in the DevOps learning journey.

\subsection{Implications to Researchers} 

Further investigation can improve the analyses about challenges and recommendations in DevOps Education. New studies can explore, for example, how difficult it is to deal with DevOps-specific challenges, how useful and practical the recommendations are to mitigate the existing challenges, and how effective the association links between challenges and recommendations are.
For example, what is the impact of the challenge \textit{"Insufficient time in the course to teach DevOps."} (P1, P3, P5, P6, P8, P9, P10, P11) on the DevOps courses? How useful is the recommendation \textit{"The assessment should be conducted with hands-on activities."} (P1, P2, P3, P5, P7)? Does the recommendation \textit{"Create tutorials to help students."} (P4, P10, P12) completely solves the challenge \textit{"Debugging lab sessions are frustrating."} (P14)? 

We collected 13 conflicts between challenges and/or recommendations. They represent inconsistencies that need to evaluated and tested in existing DevOps courses.
Researchers should pay attention to the fact that the conflicts do not necessarily need to be resolved in all cases because the characteristics of a specific course or a specific concept can decide which side of the conflict is more appropriate.

For example, considering Conflict (ix) in Section \ref{subsection:conflicts}, small examples can be helpful to teach some DevOps concepts, but others need more complex examples. The specific moment in a course could influence if small or more complex examples are preferable to teach a particular DevOps concept.

In addition to that, we also identified many challenges that have no association links with recommendations to mitigate them. These are candidates for future investigation as well. 

Although the set of challenges and recommendations identified by this work is substantial, it is not final. Researchers can use and apply other existing research strategies to gather more information and improve this dataset.

\section{Threats to validity}  \label{sec:threatsToValidity}

In this section, we discuss the threats to validity of this study in the context of qualitative research~\cite{larios:2020, guba:1981, korstjens:2018}.

\textit{Transferability}. Transferability is related to the degree to which our results can be transferred to other contexts. Our study is based on semi-structured interviews collecting the experience of 14 participants. We selected educators from three continents (North America, South America, and Europe) focused on diversity, from academia and industry. The findings of this work are robust and should be present in most DevOps courses. We concluded the interview phase only once we reached about 80\% saturation. Note that the exact saturation levels might have been different if we had interviewed participants in a different order.


\textit{Credibility}. Credibility refers to whether the research results are perfectly drawn from the original data. We adopted several strategies to ensure credibility: (a) we employ double-review in all research steps and blind-review in many of them; and (b) we discussed the findings several times between the authors of this paper to mitigate bias from the researchers involved in the study, and bias from the interviews which started with interviewees from Brazil. We also conducted a comparison with the result from Fernandes' et al.'s study~\cite{fernandes:2020}, which uses a systematic literature review as an information collection instrument. We suggest future research to validate the themes obtained.

\textit{Confirmability}. Confirmability refers the degree to which other researchers can check the findings. We do not have the participants' permission to share the transcripts of the interviews. We identified each challenge and recommendation by quoting participants as far as possible. We shared the interview script in our online research artifact~\cite{researchArtifact}.

\section{Related work}   \label{sec:relatedWork}


DevOps is an emergent research area in software engineering, and there are also a growing number of studies in the educational software engineering area~\cite{alves:2021, pang:2020}.  Recently, Fernandes et al.~\cite{fernandes:2020} conducted the first systematic literature review in DevOps Education, identifying 18 papers presenting teaching experience in DevOps courses.


Our work uses Fernandes et al.'s~\cite{fernandes:2020} study as a foundation, who also investigate the challenges and recommendations in teaching DevOps from an Engineering perspective. Different from their work, we use interviews with DevOps educators as data collection technique. Both studies adopted theme-based organization of the results, but we use thematic analysis as the qualitative data analysis method, while Fernandes et al.'s~\cite{fernandes:2020} study uses ad-hoc codification. Our interview questions are based on the themes provided by their study, and we conducted the thematic analysis using these themes as our initial ones -- although we have modified two out of the seven themes during our analysis. 

Alves et al.'s study~\cite{alves:2021} reports the experience of an experimental course with undergraduate students that combines DevOps concepts with Agile, Lean, and Open-source practices. The authors present a set of recommendations and challenges to guide DevOps educators, which came from (i) insights from professors; (ii) feedback obtained via a survey from students and partners who have already gone through the discipline; and (iii) data analysis using data mining from the course repository. Despite its contributions, Alves et al.'s study~\cite{alves:2021} is limited since it only shows the perspective of one course, at one university. We believe we have mitigated this limitation by interviewing different educators from different courses, universities, and countries, giving a more general perspective on challenges and recommendations.

Hobeck et al.'s study \cite{hobeck:2021} reports on the experience of teaching DevOps involving two universities (in the US and Germany), implementing a flipped classroom format. Similar to our work, their study discusses challenges from the instructors and students, like "constant change of DevOps tools". Also, it provides lessons learned composed of instructors' observation and students' feedback, including the recommendation to adopt an inverted classroom format to facilitate the transition to an online course format. The authors' paper shows how to mitigate the challenges. In contrast, our paper presents links from challenges to recommendations, representing a mitigation/solution proposal. We believe our paper offers more breadth since we present a set of links involving data from 14 interviews from different courses.


Using grounded theory as the qualitative research method, Pang et al.~\cite{pang:2020} study DevOps Education from academic and industrial perspectives. They also present challenges and recommendations in teaching DevOps through analyzing course syllabi from the top 50 institutions according to the 2017 QS World University Rankings by Subject - Engineering and Technology. As we interviewed the professors, we were able to obtain information about challenges and recommendations not exposed in a syllabus.

Leite et al. \cite{leite:2019} presented a survey based on a literature review. They developed a conceptual framework to guide the exploration of DevOps tools, implications, and challenges. Their study does not focus on the DevOps Education area, even though a small number of 5 challenges and 8 recommendations are education-focused. Our study identified all the educational challenges mentioned in the Leite et al. \cite{leite:2019}'work except for the challenge \textit{"Evaluating infrastructure code and non-functional concerns are harder to automate."}. Also, 5 of the 8 recommendations cited by them were not identified in our study such as \textit{"Automate the generation of assessment reports based on patterns of tasks, commits, branches, tests, and the source code itself."}, and \textit{"There are stable platforms for auto-grading coding assignments (autolab.github.io)"}. Our study shows a large set of challenges and recommendations in the DevOps education area when compared to this previous work \cite{leite:2019}.

\section{Conclusion}   \label{sec:conclusion}

DevOps education represents an urgent industry demand for software engineering courses. Our work aims to identify the main challenges and recommendations for DevOps teaching to support a better plan and execution of DevOps courses. We performed an interview study with 14 DevOps educators from different universities and countries. The study identified 83 challenges that educators encounter when planning and executing their DevOps courses.  

To improve the quality of DevOps teaching, this paper also presented 185 recommendations given by the interviewees. Some of these recommendations can support educators in addressing the identified challenges. Despite the 558 links found between the identified challenges and recommendations, there are still open challenges. For example, we found 13 conflicts between challenges and/or recommendations that highlight research opportunities on new educational methods and tools. Thus, this work contributes to both the teaching and the research area of DevOps Education. We intend to investigate the open challenges and propose new solutions and recommendations for them as future work.


\textbf{Acknowledgements.} We thank all the educators who contributed to our study. This work is partially supported by INES (www.ines.org.br), CNPq grant 465614/2014-0, CAPES grant 88887. 136410/2017-00, FACEPE grants APQ-0399-1.03/17, PRONEX APQ/ 0388-1.03/14, and IFRN.

\bibliographystyle{ACM-Reference-Format}
\bibliography{main}

\end{document}